\documentstyle[epsfig]{europhys}

\newcommand{\Label}[1]{\label{#1}}
\newcommand{\Bibitem}[1]{\bibitem{#1}}        
\newcommand{\be}{\begin{equation}}
\newcommand{\ee}{\end{equation}}
\newcommand{\ba}{\begin{eqnarray}}
\newcommand{\ea}{\end{eqnarray}}

\newcommand{\nn}{\nonumber\\}
\newcommand{\Ref}[1]{(\ref{#1})}
\newcommand{\av}[1]{\left \langle #1\right \rangle}

\newcommand{\br}{{\bf r}}
\newcommand{\bv}{{\bf v}}
\newcommand{\bk}{{\bf k}}

\newcommand{\bg}{{\bf g}}

\newcommand{\bQ}{{\bf Q}}
\newcommand{\bfe}{{\bf e}}

\newcommand{\calO}{{\cal O}}

\newcommand{\calL}{{\cal L}}
\newcommand{\calP}{{\cal P}}

\newcommand{\calG}{{\cal G}}
\newcommand{\calLbar}{\overline{\cal L}}
\newcommand{\calLtilde}{\widetilde{\cal L}}
\newcommand{\calLtildeinvert}{\widetilde{\cal L}^\epsilon}
\newcommand{\calLinvert}{\calL^\epsilon}
\newcommand{\bF}{{\bf F}}

\newcommand{\bR}{{\bf R}}

\newcommand{\bsigma}{{\sigma}}

\begin{document}
\euro{73}{2}{183-189}{2006} \Date{}

 \shorttitle{Green-Kubo formulas for
non-conservative fluids}

\title{\center {New Green-Kubo formulas for transport coefficients
in hard sphere-, Langevin fluids and the likes}}

\author{M.H. Ernst\inst{1,}\inst{2}
and R. Brito\inst{2}}

 \institute{
\inst{1}{Institute Theoretical Physics, Universiteit Utrecht,
 3508 TD Utrecht, The Netherlands}\\
 \inst{2}{Dpt.~F\'\i sica Aplicada I and GISC, Universidad
 Complutense, 28040 Madrid, Spain } }

 \pacs{ \Pacs{05}{40.-a}{Fluctuation phenomena, random processes, noise
 and Brownian motion} \Pacs{05}{20.Dd}{Kinetic theory} }

\maketitle

\begin{abstract}

We present generalized  Green-Kubo expressions for thermal
transport coefficients $\mu$ in non-conservative fluid-type
systems, of the generic form, $\mu$ $= \mu_\infty$ $+\int^\infty_0
dt V^{-1} \av{I_\epsilon \exp(t {\cal L}) I}_0$ where $\exp(t{\cal
L})$ is a pseudo-streaming operator. It consists of a sum of an
instantaneous transport coefficient $\mu_\infty$, and a time
integral over a time correlation function in a state of thermal
equilibrium between a current $I$ and its conjugate current
$I_\epsilon$. This formula with $\mu_\infty \neq 0$ and
$I_\epsilon \neq I$ covers vastly different systems, such as
strongly repulsive elastic interactions in hard sphere fluids,
weakly interacting Langevin fluids with dissipative and stochastic
interactions satisfying detailed balance conditions, and "the
likes", defined in the text. For conservative systems the results
reduce to the standard formulas.
\end{abstract}

The Green-Kubo formulas for thermal transport coefficients in
simple classical fluids with smooth {\it conservative}
interactions are widely used, and generally accepted
\cite{Hansen-McDonald,Allen-Tildesley,ED75} as exact expressions
for general densities, as long as the deviations from equilibrium
and the gradients are small, and the transport coefficients exist.
This standard expression is given in terms of an equilibrium time
correlation function between $N$-particle currents $I$ and
$I_\epsilon$, i.e.
\be\Label{1}
\mu = \int_0^\infty \! dt \lim_{V\to\infty} \frac{1}{V} \langle
I(0)I(t)\rangle_0,
\ee
where $\mu$ denotes a typical transport coefficient, $\langle
\dots \rangle_0$ is an average over a thermal equilibrium ensemble
at temperature $T=1/k_B\beta$, with a distribution function
$\rho_0\sim \exp[-\beta H]$, and $\lim_{V\to\infty}$ denotes the
thermodynamic limit. In the sequel this limit is understood, but
not explicitly written. For convenience, time is taken to be
continuous. The time evolution of a dynamical variable, $I(t)=
e^{t\calL} I(0)$, can be described by a streaming operator $
e^{t\calL}$, which generates a time evolution that is invariant
under the time reversal transformation.  Consequently, the
Liouville operator satisfies, $\calLinvert = -\calL$, where
$\calLinvert$ denotes the time reversal transform of $\calL$. As
$\calL$ and the currents $I$ contain in general interparticle
forces, the standard Green-Kubo formula \Ref{1} is ill-defined if the
interaction potentials are not sufficiently smooth, and it does
neither apply to hard sphere or hard core interaction potentials
(non-conservative impulsive forces), nor to dissipative and
stochastic forces, which are not derivable from an interaction
potential.

{\it Problems addressed:} The goal of  this letter is to present
generalized Green-Kubo expressions, without any restrictions to
dilute systems, for a large class of complex (non-conservative)
fluids, i.e. $N$-particle systems, that do approach a state of
thermal equilibrium, but whose equations of motion  have
ill-defined forces, or/and lack time reversal invariance. The
forces may be impulsive, dissipative, stochastic, and possibly
include conservative forces as well. The generic form of this
generalized Green-Kubo formula is,
\be\Label{2} \mu=\mu_\infty
+V^{-1}\int_0^\infty\! dt
 \langle I_\epsilon e^{t\calL}I\rangle_0,
\ee
where it must be possible to represent the time evolution through
a  {\it pseudo}-streaming operator $\exp(t\calL)$, which is
well-defined for {\it all} points in the relevant phase space, and
generates the proper trajectories of the dynamical variables when
used {\it inside statistical averages} $\av{\cdots}_0$. The prefix
{\it pseudo} refers to the fact that the representation
$\exp[t\calL]$ is only valid inside averages. An assessment of the
type of systems to which this Green-Kubo formula  is also
applicable, called "the likes", and an outline of the derivation
and implications are the main goals of the present article.
 The discussion  covers
such vastly different systems as microscopic fluid models with
strongly repulsive elastic interactions (hard spheres, square well
potentials) for which the interparticle pseudo-forces and the
pseudo-Liouville operators in Eq.\Ref{2} are well-defined, as well
as mesoscopic weakly interacting Langevin fluids with dissipative
and stochastic forces.

 There seem to exist in the literature only three cases
where explicit Green-Kubo formulas of the generalized form \Ref{2}
have been derived for transport coefficients in complex fluids,
each referring to a very different system and to a very different
transport coefficient, obtained by very different approaches. The
first case in Ref.\cite{Pias} refers to Brownian dynamics. Here
the friction coefficient of a Brownian particle in a fluid, both
modeled by hard spheres,  is derived with the help of a delicate
analysis, using an asymptotic expansion in the small ratio $m/M$
of the hard sphere mass over the mass of the Brownian particle. In
this limit the interactions between the heavy particle and the
liquid particles becomes weak, and the kinetic equation for the
distribution function of the heavy particle reduces to a Fokker
Planck equation. Here the friction coefficient is expressed as a
generalized Green-Kubo formula \Ref{2}, and contains a time
correlation function between a pseudo-interparticle force and its
conjugate. A possible extension of this method to complex fluids
is not obvious.

A second case, also concerning hard spheres, can be found in Ref.
\cite{DE04} for the shear viscosity of an elastic hard sphere
fluid, derived by a transformation of the corresponding
Einstein-Helfand formula, or by a limiting procedure applied to
the standard Green-Kubo formulas for soft spheres with a repulsive
interaction potential $V(r)\sim 1/r^n$ with $n\to \infty$. The
limiting procedure cannot be extended to complex fluids, but the
first route via an Einstein-Helfand formula is possible, following
Ref. \cite{DE04}.

A third case concerns a Langevin fluid \cite{PE-Vasquez}, where
standard linear response theory \cite{Hansen-McDonald} is used to
obtain the long time diffusion coefficient of a colloidal
suspension in the generic form \Ref{2}. This is done by modeling the
short time diffusion coefficient by a phenomenological relaxation
constant, used as input in the Langevin equation. The extension of
that method for a weakly interacting Langevin fluid to strongly
interacting hard sphere systems does not seem feasible either.

The question of interest in this article is then: do these
scattered results  have something in common that makes the
structure \Ref{2} generic for a larger class of physical systems?
The answer to this question will be  given below.

{\it Properties of $\calL$:} Suppose that $\calL$ has been
constructed. As we are studying equilibrium time correlation
functions, we have to impose the condition of stationarity,
\be\Label{3}
\langle A(0)B(t)\rangle_0= \langle A(-t)B(0)\rangle_0.
\ee
 Here the time evolution of $A(-t)\equiv \exp(t\calL^\epsilon)
A(0)$ (where $t>0$) is generated by the backward or time reversed
streaming operator. Defining the transpose $\calLtilde$ through
the relation, $\int d\Gamma A e^{t\calL} B= \int d\Gamma B
e^{t\calLtilde} A$, we find by comparing integrands in Eq. \Ref{3}
the necessary condition on $\calL$ in the form of a commutation
relation, i.e.
\be\Label{4}
\rho_0\calL =\calLtildeinvert\rho_0\ \ \ \mbox{or\ } \ \ \
\calLbar=\calLtildeinvert,
\ee
where $\calLbar$ is defined through $\rho_0\calL\equiv \calLbar
\rho_0$. The above ingredients are sufficient to derive Green-Kubo
type formulas for the Navier-Stokes transport coefficients using
linear response theory. The class of "the likes" consists of those
microscopic or mesoscopic models that have: (i) a pseudo-streaming
operator of the form $\exp[t \calL ] $, (ii) a stationary
distribution given by the thermal $\rho_0 $, where (iii)  $\rho_0
$ and  $\calL $ satisfy the commutation relation \Ref{4}. Note that
for conservative interactions, where
$\calLtilde=-\calL=\calL^\epsilon$, the condition \Ref{4} is
trivially satisfied as $\calLbar =  \calLtildeinvert =\calL$, and
$\calL$ and $\rho_0$ commute.

There exist two important simple realizations of the previous
scenario. The first and most important class are the fully
microscopic models of hard sphere fluids, which are prototypical
for the strongly repulsive hard core interactions of classical
fluids. The other class are mesoscopic Langevin fluids with weak,
dissipative and stochastic forces, where the interactions are
modeled by phenomenological input parameters.

For hard spheres the standard Liouville operator for conservative
systems is ill-defined because of the impulsive interaction
forces. For this case the construction of pseudo-Liouville
operators $\calL$ in terms of binary collision operators is far
from trivial \cite{EDHvL}. The resulting expressions for $\calL $
in terms of binary collision operators satisfy the commutation
relation \Ref{4}, as shown in \cite{EDHvL}. This pseudo-Liouville
operator is not odd in the velocities. So $ \calL^\epsilon \neq -
\calL$, and consequently not all trajectories generated by $\exp(t
\calL)$ are time reversal invariant.

For Langevin fluids the construction of the pseudo-Liouville
operator $\calL$ is rather trivial, as this operator can be simply
obtained  from the corresponding Fokker-Planck equation,
$\partial_t \rho=\calLtilde\rho$, for the $N$-particle probability
distribution $\rho(\Gamma,t)$ \cite{Risken}. As this equation
satisfies the detailed balance condition,  the stationary solution
of $\partial_t \rho_0 = \calLtilde\rho_0 = 0$ is the thermal
distribution $\rho_0$. Also here it is sufficient to define
pseudo-streaming operators $e^{t\calL}$ for the time evolution of
dynamical variables only inside averages. This may be done through
the relation,
\be\Label{5} \langle A(t)\rangle
=\int d\Gamma A e^{t\calLtilde} \rho(\Gamma,0)= \int d\Gamma
\rho(\Gamma, 0) e^{t\calL} A = \langle e^{t\calL}A\rangle.
\ee
Here the pseudo-Liouville operator $\calL$ is the transpose of the
Fokker-Planck operator. In this formulation $\calL$ is not the
infinitesimal generator of the Langevin equations, but an
effective operator, that acts inside averages, in which the rapid
fluctuations of the random forces have been averaged out. Because
of the presence of dissipative forces, the Fokker-Planck equation
is not invariant under the transformation of time reversal.
Consequently, $\calLinvert\neq -\calL$. For Fokker-Planck
equations, satisfying the detailed balance criteria, the operator
$\calL$ satisfies the commutation relation \Ref{4} (see
\cite{Risken}).

{\it Linear response theory:} Having discussed the properties of
the pseudo-Liouville operator, one can apply  linear response
theory for calculating thermal transport coefficients in fluids
\cite{Hansen-McDonald,ED75}. Consider, for simplicity, a system
with a single conserved density $e(\br)$. Extensions to more
conservation laws are straightforward. Let $e_\bk$ be the Fourier
mode of the conserved density, $ e_\bk = \sum_i \delta\epsilon_i
\exp[-i \bk \cdot \br_i]$ with fluctuation $ \delta\epsilon_i=
\epsilon- \av{\epsilon}_0$.  The microscopic equation of motion
for the Laplace transform $e_{\bk z}$ of $e_\bk (t) $ is
$(z-\calL) e_{\bk z} = e_\bk$. In the {\it long wave length} limit
$( k \to 0)$ the hydrodynamic propagator satisfies the relation,
\be \Label{6} G(\bk,z) \equiv
\av{e_\bk|e_{\bk z}} \simeq
 \av{e_{\bf 0}|e_{\bf 0}}/[z+k^2 D],
\ee
where $D= \mu / \av{e_{\bf 0}|e_{\bf 0}}$ is the diffusivity, and
$\mu$ the kinetic coefficient. The standard projection operator
method \cite{Hansen-McDonald} enables us to derive in a few lines
an exact expression for $D$ or $ \mu$. To do so we introduce the
projection operator $\calP = 1- \calP_\bot$ acting on a dynamic
variable $b_\bk$ as $\calP b_\bk = e_\bk
{\av{e_\bk|b_\bk}}/{\av{e_\bk|e_\bk}}$. The inner product is
defined as a thermal average, $\av{a_\bk|b_\bk} = V^{-1}
\av{a^*_\bk b_\bk}_0$. Application of this method yields,
\be \Label{7} [z-\calP\calL\calP
-\calP\calL\calP_\perp(z-
\calP_\perp\calL\calP_\perp)^{-1}\calP_\perp\calL\calP] \calP
e_{\bk z}= e_\bk.
\ee
Writing this in component form, and comparing the result with
\Ref{6} yields for $D$,
\be \Label{8}
D =  - \calL im \;\Re \left[ {\av{e_\bk|\calL e_\bk}} +
\av{\calL^\epsilon e_\bk | \calP_\perp \hat{\calG}_z \calP_\perp
\calL e_\bk} \right] /[ k^2  \av{e_\bk|e_\bk}],
\ee
where $\calL im$ represents the double limit, $\lim_{z\to
0}\lim_{k\to 0}$, $\Re$ denotes the real part, and $\hat{\calG}_z$
reduces in  the small-$k$ limit to the projected resolvent$,
\calP_\perp (z-\calL)^{-1}\calP_\perp$. To obtain the second term
we have used the relation $\av{e_\bk| \calL \calP_\bot \cdots} =$
$\av{\calL^\epsilon e_\bk| \calP_\bot \cdots }$, based on the
commutation relation \Ref{4}. This term can be expressed as a
current-current correlation function.

Consider the so called Euler matrix $\av{e_\bk|\calL e_\bk} =
-ik\av{e_\bk|j_\bk}$. Its imaginary part would yield the Euler
equations for a fluid with the standard conservation laws. As
conserved densities have in general a definite parity in the
velocities (mass, momentum, energy), the Euler matrix element
would vanish for conservative systems, where $\calLinvert=-\calL$.
However, for systems whose pseudo-streaming operator lacks time
reversal invariance, $\calLinvert\neq -\calL$, and the matrix
element has a non-vanishing real part for small $k$ of ${\cal O}
(k^2)$, i.e.
\be\Label{9}
\mu_\infty=\av{e_{\bf 0}|e_{\bf 0}}D_\infty= -
\lim_{k\to\infty}k^{-2}\Re \av{e_\bk|\calL e_\bk}.
\ee
Next consider the second term in Eq.~\Ref{8}. Here the total
microscopic flux, $J=\lim_{k\to 0} j_\bk$, is related to the
Fourier mode $e_\bk$ of the conserved density through the local
conservation law, $\partial_t e_\bk = \calL e_\bk=-ikj_\bk$.
Similarly, we define the conjugate current, $J_\epsilon=\lim_{k\to
0} j^*_{\epsilon\bk}$ through $\calLinvert
a_{\bk}^*=-ikj^*_{\epsilon\bk}$. Inserting these definitions into
Eq.~\Ref{8} yields finally for the kinetic coefficient $\mu$ the
generalized Green-Kubo formula \Ref{2}, where $I$ and $I_\epsilon$
are subtracted currents, defined as $I=\calP_\perp J$. This
completes the formal derivation of Eq.~\Ref{2}, and it remains to
work out the explicit expressions for $I$, $I_\epsilon$ and
$\mu_\infty$.  In the remaining part of this letter we present
some applications to a heat conducting random solid, an isothermal
Langevin fluid, and a hard sphere fluid.

{\em Heat conducting random solid:} A realization of a system in the previous
section is a quenched Langevin fluid with heat conduction.
It consists of $N$ point particles, {\it quenched} in a  random
configuration $X = \{\br_i |i=1,\cdots, N\}$. Each particle is
characterized by its internal energy $\epsilon_i$, collectively
denoted by $\bfe=\{\epsilon_i (t) |i=1,\cdots, N\}$, which are the
only dynamical variables in the model. The internal energy
corresponds  to many internal degrees of freedom, described by a
density of states $\sim \epsilon^\alpha$, where $\alpha$ is
proportional to the number of internal degrees of freedom $(\alpha
\gg 1)$. Energy is exchanged between the particles through
dissipative and stochastic forces of finite range, which obey
detailed balance criteria. The total energy, $E=\sum_i
\epsilon_i(t)$ is conserved, and the $N$-particle system
approaches a state of thermal equilibrium. The system supports a
local heat current, driven by gradients in the local energy
density or temperature field. So, at the macroscopic level
Fourier's law of heat conduction applies, at least if the density
of particles is above a critical percolation threshold. Our goal
is to derive a Green-Kubo expression for its heat conductivity.

The Langevin equations for the time evolution of the dynamical
variables  $\epsilon_i (t)$ and the corresponding Fokker-Planck
equation for the $N$-particle probability density $\rho(\bfe, t|X)$
in a fixed configuration $X$ of $N$ point particles have been
derived to dominant order $\calO (1/\alpha)$ in Ref.
\cite{REE98,RE05}. We only quote the adjoint  Fokker-Planck
operator $\calL$ which reads,
\be \Label{10}
\calL =
\textstyle{\sum_{i<j}} \lambda_{ij} \left[ (\epsilon_j-\epsilon_i)+
\alpha^{-1}  \epsilon_i\epsilon_j
\partial_{ij} \right] \partial_{ij} =\calLinvert.
\ee
Here $\partial_{ij} =\partial/\partial\epsilon_i -
\partial/\partial\epsilon_j$, and $\lambda_{ij} =
\lambda_0 w(r_{ij})$ with a positive range function $w(r_{ij})$
vanishing for $r_{ij}\geq r_c$, and $\gamma_0$ is a
phenomenological model parameter. The equation above shows that
$\calLinvert=\calL$ because $\epsilon_i$ does {\it not} change
sign under a time reversal transformation. The microscopic energy
fluxes for small $k$ are obtained as \cite{RE05},
\ba\Label{11}
\calL e_\bk  & \simeq & -i\bk \cdot
\textstyle{\sum_{i<j}} \lambda_{ij} (\epsilon_j-\epsilon_i) \br_{ij}
\equiv -i\bk\cdot \bQ \nonumber \\
\calLinvert e^*_\bk & \simeq & -i\bk\cdot \bQ_\epsilon = \calL
e^*_\bk =+i\bk\cdot \bQ ,
\ea
and there are no subtracted parts. Note the relation $
\bQ_\epsilon = -\bQ$. The Green-Kubo formula for heat conductivity
in the   random solid is then $\lambda=\lambda_\infty+\lambda^{dd}$, with
\be\Label{12}
\lambda^{dd}  = 
(\beta/dTV) \int_0^\infty dt \av{\bQ_\epsilon
\cdot e^{t \calL} \bQ}_0
=  -(\beta/dTV) \int_0^\infty dt \av{\bQ \cdot
e^{t\calL} \bQ}_0.
\ee
The Green-Kubo expression \Ref{12} has the generic form for
systems with dynamics lacking time reversal invariance.  As this
formula can also be derived from an Einstein-Helfand formula,
 $\lambda$ is necessarily  positive. Because $\calLinvert\neq
-\calL$, necessarily $\lambda_\infty\neq 0$.  The instantaneous
heat conductivity  is equal to the mean field approximation to the
heat conductivity, calculated in Refs.\cite{REE98,RE05}, i.e. $
\lambda_\infty = (\lambda_0 n^2 C_v/2d) [w R^2] $ where $ [w R^2]=
\int d\bR w(\bR) R^2$ and $C_v$ is the specific heat per particle.
In fact, the remaining time integral in \Ref{12} gives at large
densities only a negligible contribution, but at the percolation
threshold both terms cancel \cite{RE05}. If the {\it standard}
Green-Kubo formula \Ref{1} would have applied, then $ \lambda_\infty =
0$, and the sign in front of the second time integral would have
been a plus sign. So, the resulting expression would have been
quite different from Eq.~\Ref{12}.

\indent{\it Isothermal DPD fluid:} Next we will consider the so
called isothermal DPD (dissipative particle dynamics)  fluid,
introduced in Ref.\cite{DPD-original}, and studied analytically
\cite{Esp-Warren,Marsh} and by means of computer simulations
\cite{Ign-EPL}. It is a mesoscopic version of a classical fluid
where the fast microscopic length and time scales are averaged
out, and DPD particles, described by their position $\br_i$ and
velocity $\bv_i$, can be thought of as soft lumps of fluid. Their
equations of motion, $ d{\bf r}_i/dt = \bv_i$ and $ m d\bv_i/dt
=\sum_j (\bF^c_{ij}+ \bF^d_{ij}+ \bF^r_{ij})$, contain
conservative forces, ${\bf F}^c_{ij} = -\partial
V(r_{ij})/\partial \br_{ij}$, as well as dissipative $(d)$ and
random $(r)$ forces, where $\bF^d $ is proportional to a
phenomenological relaxation parameter $ \gamma_0$. Explicit
expressions for  $\bF^d $ and $\bF^r $ can be found in
Refs.~\cite{Esp-Warren,Marsh}. All forces have a finite mesoscopic
range $r_c$. The corresponding Fokker-Planck equation satisfies
the detailed balance condition. The pseudo-Liouville operator,
$\calL = \calL_c +\calL_d$, is obtained as the transpose of the
Fokker-Planck operator. Its conservative part $\calL_c$ includes
the inertial term and the conservative forces (where $
\calL_c^\epsilon = - \calL_c $). Its dissipative part $\calL_d$
contains the combined action of the dissipative and stochastic
forces (where $ \calL_d^\epsilon = \calL_d $). Consequently the
instantaneous transport coefficient depends solely on the
dissipative force. In the DPD fluid total mass and momentum are
conserved, but {\it not} the total energy. So mass density $n_\bk$
and momentum density $\bg_\bk$ are the only slow Fourier modes.
Therefore, the fluid exhibits shear and bulk viscosity. The
transport coefficients are given by the matrix generalization of
Eq.\Ref{8} (see \cite{ED75}). Here we restrict ourselves to the
shear viscosity $\eta$. It is given by an expression similar to
\Ref{8} with $e_\bk$ replaced by the transverse momentum density
$g_{\bk \bot} = \sum_i m v_{iy} \exp[-ik r_{ix}]$ where $\bk$ is
taken parallel to the $x$-axis for convenience, and it is straight
forward to obtain,
\ba \Label{13}
\eta &=& \eta_\infty + (\beta/V)\int^\infty_0 dt
\av{(J^c_{xy}-J^d_{xy}) e^{t \calL} (J^c_{xy}+J^d_{xy})}_0 \nn
&=& \eta_\infty +\eta^{cc} + \eta^{cd} +\eta^{dc} +\eta^{dd}
\ea
For {\it small}  $k$ the relation $\calL g_{\bk \bot} \simeq
-ikJ_{xy}$ has been used, where the total momentum current $J_{xy}
= J^c_{xy}+J^d_{xy}$, and $J_{\epsilon xy} = J^c_{xy} - J^d_{xy}$.
Here $J^c_{xy}$ is the momentum current in the standard Green-Kubo
formula for conservative forces, given here by $\eta^{cc}$, and
$J^d_{xy}=-J^d_{\epsilon,xy} = \sum_{i<j} r_{ij,x} F^d_{ij,y}$ is
the dissipative current.  The instantaneous viscosity
$\eta_\infty$, which is proportional to $k^{-2} \av{g_{\bk
\bot}|\calL g_{\bk \bot}}$, is equal to the quantity $\eta_D = m
n^2\gamma_0 [g_0 w R^2]/2d(d+2)$, where $g_0(r)$ is the radial
distribution function in thermal equilibrium, calculated in
Ref.\cite{DPD-original,Marsh} as the mean field approximation  to
the viscosity. Also note that the commutation relation \Ref{4}
implies that $\eta^{cd}=\eta^{dc}$, which contributions are in
general non-vanishing. If the {\it standard} Green-Kubo formula
would apply, then $ \eta_\infty=0$, and the minus sign in the
integrand would be a plus sign.

\indent{\it Hard sphere fluid:} We start from the expressions for
$ \calL = L_+$ and $\calL^\epsilon = - L_-$ in the notation of
hard sphere kinetic theory  \cite{EDHvL,vBE-JSP,dSEC,DE04}, i.e.
\be \Label{14}
L_\pm =L_0 \pm \textstyle{\sum_{i<j}} T_\pm (ij); \qquad
\bar{L}_\pm =L_0 \pm
\textstyle{\sum_{i<j}} \overline{T}_\pm (ij).
\ee
Here $L_0 = \sum_i \bv_i \cdot \partial/\partial \br_i$ is the
inertial part. The binary collision operators satisfy the
relations $\overline{T}_\pm = \widetilde{T}^\epsilon_\pm$ and
${T}^\epsilon_\pm =T_\mp$, and their explicit forms can be found
in the literature \cite{EDHvL}. Then the linear response theory
leads to the generalized Green-Kubo formulas \Ref{2} for the
Navier-Stokes transport coefficients in hard sphere fluids, which
reads e.g. for the shear viscosity,
\be \Label{15}
 \eta = \eta_\infty
+({\beta}/{V})\int^\infty_0\! dt\, \av{J_{-xy} e^{tL_+}J_{+xy} }_0.
\ee
Here  $J_{xy}$ is the total momentum current, obtained for small
$k$ from the relation $L_\pm g_{\bk \bot} \simeq - ik J_{\pm xy} =
-ik (J^k_{xy} + J^v_{\pm xy}) $,  where $(k)$ refers to the
kinetic flux, $J^k_{xy} = \sum_i mv_{ix}v_{iy}$, and $(v)$ to the
collisional transfer flux,
\ba \Label{16}
J^v_{\pm xy} &=& \pm \textstyle{ \sum_{i<j}} T_\pm(ij) m(v_{iy}
r_{ix} + v_{jy} r_{jx})
\nn &=& \textstyle{\sum_{i<j}} m \sigma^d \int^{(\mp)} d \hat{\bsigma}
(\bv_{ij} \cdot \hat{\sigma})^2 \hat{\sigma}_x \hat{\sigma}_y
\delta (\br_{ij} - \sigma \hat{\sigma}).
\ea
The constraint $(\mp)$ on the $\hat{\sigma}$-integration restricts
that integral to the pre/post-collision hemisphere where $\mp
\bv_{ij} \cdot \hat{\sigma} \geq 0$. The explicit expressions for
the currents, which are identical to those in Ref.\cite{DE04}, are
very different from the ones appearing in the standard form
\Ref{1}, and also very different from those in those in the
Langevin fluids. The instantaneous viscosity $\eta_\infty$ is
proportional to $k^{-2} \av{g_{\bk \bot}|\calL g_{\bk \bot}}$, and
is calculated in Ref. \cite{DE04} as $\eta_\infty  = \rho\sigma^2/
[d(d+2)t_E]$, where $\sigma$ is the diameter of the hard spheres,
and $t_E$ the Enskog mean free time. The same instantaneous
contributions can also be identified \cite{vBE-JSP,EB05-PRE} in
the Enskog theory \cite{Chapman-Cowling} for the hard sphere
fluid. Details are presented in \cite{EB05-PRE}.

We conclude with a number of comments:\\
(i) We have derived generalized Green-Kubo formulas, Eq. \Ref{2},
for non-conservative systems, ranging from microscopic hard sphere
fluids with elastic impulsive interactions, to mesoscopic Langevin
fluids with weak dissipative and stochastic interactions. Moreover
a fundamental commutation relation \Ref{4} has been formulated
which guarantees that the time correlation functions are
stationary, and  characterizes those systems for which  the
transport coefficients are given by the generalized form \Ref{2}.
For conservative systems the generalized formulas reduce to the
standard Green-Kubo expressions \Ref{1}.

(ii)  The generalized formula \Ref{2} does not apply to fluids of
inelastic hard spheres, which are models for granular fluids,
although infinitesimal generators of the form \Ref{14} have been
constructed. The reason is that such dissipative systems, either
freely cooling or driven by an energy source, do not reach a
stationary state described by the thermal distribution $\rho_0 $.

(iii)  From the existence of pseudo-streaming operators for hard
sphere fluids an interesting side result can be deduced. It allows
us to introduce pseudo-interparticle forces \cite{Pias,DE04}, $
{\bf F}^{ps}_{ij}$ and conjugate forces through the relation $
\calL m \bv_i = \sum_{j (\neq i)}{\bf F}^{ps}_{ij} $, and   $
\calL^\epsilon m \bv_i = \sum_{j (\neq i)}{\bf
F}^{ps}_{\epsilon,ij} $ respectively. An identity, derived in
\cite{DE04}, states that $ \av{{\bf F}^{ps}_{\epsilon,ij,x}(0){\bf
F}^{ps}_{ij,y}(t)}_0  = {\cal C}\delta_{xy} \delta (t)$, where
${\cal C}$ is given explicitly. Comparison with the time
correlation function for the Langevin forces suggests that the
pseudo-interparticle forces in hard sphere systems can be
considered as white noise.

(iv) Another side result of interest, following from \Ref{14}, is
the observation that the frequently cited Green-Kubo formulas for
the DPD fluid \cite{PE95}, regarding for the viscosity for instance,
$\eta=\eta^{cc}-\eta^{dd}$, are not correct (compare with (13)).
Equally
incorrect is the standard Green-Kubo formula \Ref{1} for the heat
conductivity of the random solid, quoted in
Ref.\cite{RE05} as $\lambda=-\lambda^{dd}$ (compare with (12)).
It was mentioned there as a possible alternative for the kinetic theory
analysis, presented in that paper. Regarding the main interest of
this letter, i.e. the generic structure of the Green-Kubo formulas
for non-conservative systems, the last comment is only of minor
importance.

\vspace{5mm} Acknowledgments: The authors thank J.~Piasecki and
P.~Espa\~nol for bringing Refs.\cite{Pias} and \cite{PE-Vasquez}
respectively to their attention. M.H.E. is supported by
Secretar\'\i a de Estado de Educaci\'on y Universidades (Spain),
and R.B. by the Universidad Complutense (Profesores en el
Extranjero). This work is financed by the research project
FIS2004-271 (Spain).

\end{document}